\documentclass[aps,prl,twocolumn,showpacs,superscriptaddress,reprint,longbibliography]{revtex4-1}

\usepackage{amsmath, amsfonts, amssymb}
\usepackage{enumerate}
\usepackage{graphicx}
\usepackage{color}
\usepackage[breaklinks]{hyperref}
\usepackage[hyphenbreaks]{breakurl}
\usepackage{paralist}

\newcommand{\eref}[1]{Eq.~(\ref{#1})}
\newcommand{\fref}[1]{Fig.~\ref{#1}}
\newcommand{\tref}[1]{Table~\ref{#1}}

\newcommand{\up}{\uparrow}
\newcommand{\dw}{\downarrow}

\newcommand{\si}{\hat{\sigma}_{0}}
\newcommand{\sx}{\hat{\sigma}_{1}}

\newcommand{\sz}{\hat{\sigma}_{3}}

\newcommand{\ti}{\hat{\tau}_{0}}
\newcommand{\tx}{\hat{\tau}_{1}}

\newcommand{\tz}{\hat{\tau}_{3}}

\newcommand{\mean}[1]{\langle #1 \rangle}

\newcommand{\mbf}[1]{\mathbf{ #1 }}

\DeclareMathOperator{\real}{Re}

\DeclareMathOperator{\sgn}{sgn}

\begin{document}


\title{All-electrical generation and control of odd-frequency $s$-wave Cooper pairs in double quantum dots}

\newcommand{\nagoya}{Department of Applied Physics, Nagoya University, Nagoya 464-8603, Japan}
\newcommand{\sapporo}{Department of Applied Physics, Hokkaido University, Sapporo 060-8628, Japan}

\author{Pablo Burset}
\affiliation{\nagoya}

\author{Bo Lu}
\affiliation{\nagoya}

\author{Hiromi Ebisu}
\affiliation{\nagoya}
 
\author{Yasuhiro Asano}
\affiliation{\sapporo}
 
\author{Yukio Tanaka}
\affiliation{\nagoya}
 
\date{\today}



\begin{abstract}
We propose an all-electrical experimental setup to detect and manipulate the amplitude of odd-frequency pairing in a double quantum dot. 
Odd-frequency pair amplitude is induced from the breakdown of orbital symmetry when Cooper pairs are injected in the double dot with electrons in different dots. 
When the dot levels are aligned with the Fermi energy, i.e., on resonance, nonlocal Andreev processes are directly connected to the presence of odd-frequency pairing. Therefore, their amplitude can be manipulated by tuning the level positions. 
Detection of nonlocal Andreev processes by conductance measurements contributes a direct proof of the existence of odd-frequency pair amplitude and is available using current experimental techniques. 
\end{abstract}

\maketitle

\emph{Introduction.---}
The symmetry analysis of Cooper pairs is a key element in the study of superconductivity. 
For example, Cooper pairs at conventional BCS superconductors form a spin-singlet even-parity state, where the electrons have opposite spins and are coupled in momentum space by the isotropic $s$-wave channel. 
A current trend in the study of superconductivity is to \textit{engineer} unconventional superconductors by breaking down symmetries of a BCS superconductor. Consequently, a new type of pairing emerges which is odd in frequency, i.e., odd under an exchange of the time coordinates \cite{Berezinskii_1974,*Berezinskii_1974R,Balatsky_1992,Tanaka_JPSJ,Eschrig_RPP}. 
Plenty of theoretical studies suggest ubiquitous presence of odd-frequency superconductivity at inhomogeneous superconducting systems \cite{Tanaka_2007,*Tanaka_2007c,Yokoyama_2012,Black-Schaffer_2012,Black-Schaffer_2013,Black-Schaffer_2013b,Sothmann_2014,Ebisu_2015,Crepin_2015,Burset_2015,Asano_2015,Mizushima_2015}. 
Unfortunately, experimental evidence for odd-frequency pair amplitude is very limited. Odd-frequency spin-triplet $s$-wave superconductivity can explain the long-range proximity effect \cite{Bergeret_2001,*Bergeret_RMP,Birge_2010}, the intrinsic paramagnetic Meissner effect \cite{Tanaka_2005,Yokoyama_2011,Asano_2011,Robinson_2015}, and the subgap structure \cite{Robinson_2015b} observed in ferromagnet-superconductor hybrids. 
However, odd-frequency pairs are mixed with conventional ones and their amplitude is not tunable but accidentally determined by the configuration of magnetic moments realized at the junction. 
To unambiguously establish the presence of odd-frequency pairing, new proposals that filter odd-frequency pairs and allow to control their amplitude are required. 


\begin{figure}[ht!]
	\includegraphics[width=1.\columnwidth]{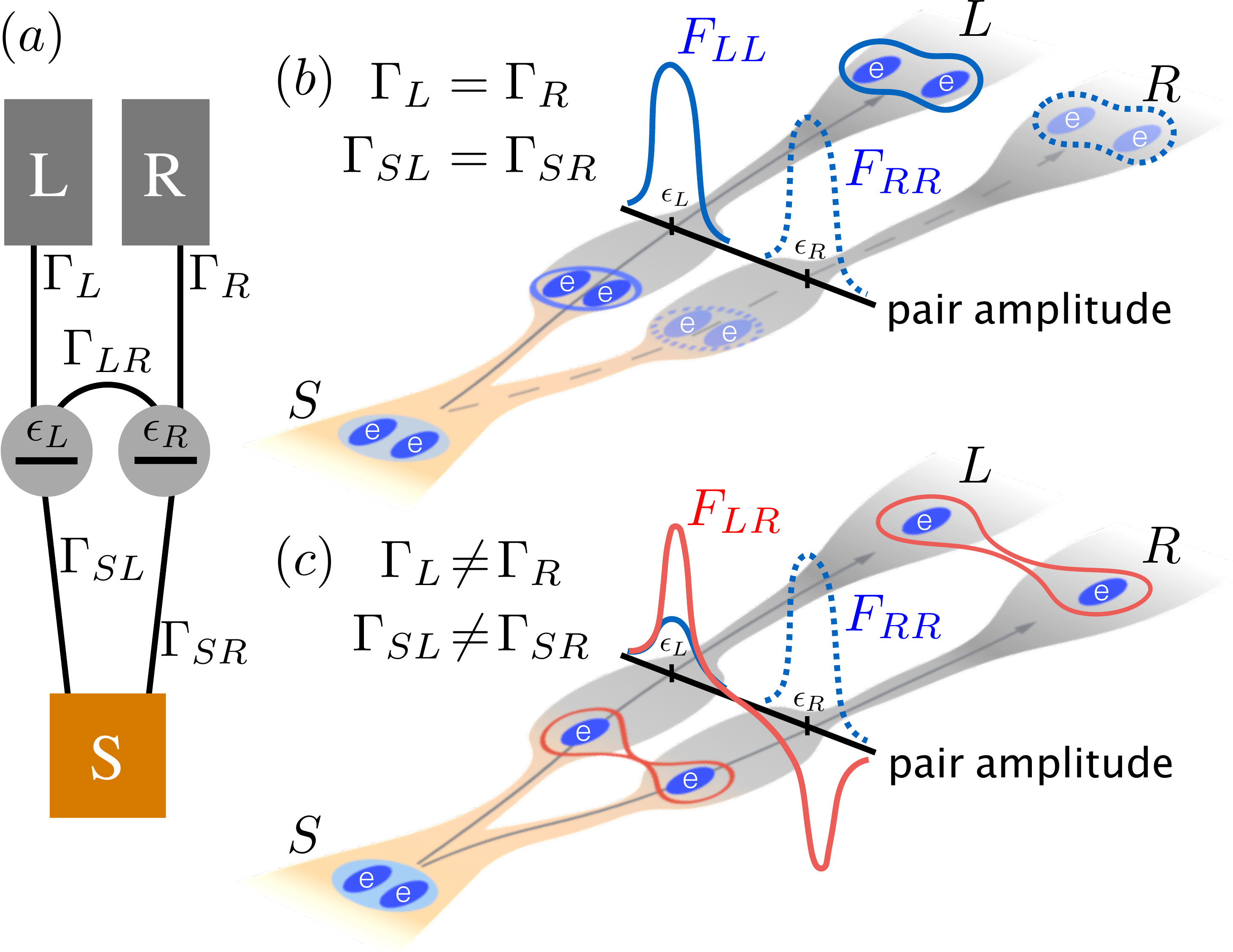}
	\caption{\label{fig:figure_sketch}
		Proximity-induced superconductivity in a DQD three-terminal device. 
		(a) Schematics of a DQD with level positions $\epsilon_{L}$ and $\epsilon_{R}$ contacted by a superconducting lead S and two normal leads L and R. 
		(b) In a local Andreev process, the electrons of a Cooper pair tunnel through one dot into the normal lead. The amplitude for these processes, $F_{LL,RR}$ (blue dashed lines), is enhanced for symmetric devices, i.e., those where each dot is similarly coupled to the leads. 
		(c) In a nonlocal Andreev process, each electron of the Cooper pair tunnels to a different lead. The nonlocal amplitude, $F_{LR}$ (red solid line), is enhanced in asymmetric devices. On resonance ($\epsilon_{L}\!=\!\epsilon_{R}\!=\!0$), $F_{LR}$ is odd in frequency if S is a BCS superconductor. }
\end{figure}


In this Letter, based on recent experimental progress on double quantum dot (DQD) Cooper pair splitter devices \cite{Hofstetter_2010,Takis_2010}, we propose a setup that allows for the detection and manipulation of odd-frequency pairing without using any magnetic elements. 
Such pair splitters consist of a DQD independently connected to two normal leads and one superconducting electrode as sketched in \fref{fig:figure_sketch}(a). 
We show that the symmetry of the induced pairing from the superconducting lead is broken due to the DQD orbital degree of freedom, thus becoming a superposition of symmetric and antisymmetric orbital states. 
For a spin-singlet superconductor, induced Cooper pairs that are antisymmetric (symmetric) in DQD space must be odd (even) in frequency according to Fermi-Dirac statistics, since parity and spin rotation symmetries are not broken. 
Cooper pairs transmitted to the same lead are in an even-frequency state [see \fref{fig:figure_sketch}(b)]. 
On the other hand, when the levels are on resonance, only odd-frequency pair amplitude is responsible for the splitting of Cooper pairs into different leads [see \fref{fig:figure_sketch}(c)]. 
Antisymmetric odd-frequency singlet pairing is greatly enhanced if each dot is coupled differently to the leads, resulting in a measurable contribution to the conductance. 
Such a connection between symmetry and microscopic transport processes is a unique feature of our proposal. 
Additionally, the amplitude of odd-frequency Cooper pairs can be controlled by tunning the DQD level positions on- or off-resonance. 
Alternatively, using a spin-triplet superconductor, the same geometry can be used for the study of Majorana edge states \cite{Nakosai_2013}. In such a case, odd-frequency triplet pairs are now transmitted to the same electrode. 
Our proposal opens a novel direction in the study of exotic Cooper pairing owing to the unique connection between symmetry of the Cooper pair and tunneling processes and due to the tunability of the pair amplitude. 

\emph{Model.---} 
DQD-based three terminal devices \cite{Recher_2001,Martin_2002,Bena_2002,Buttiker_2003} are an ideal platform for exploring the symmetry of induced pairing. 
Recent experiments are very well modeled by two-level systems and show an exquisite degree of tunability \cite{Takis_2010,Hofstetter_2010,Burset_2012b,Csonka_2014,Csonka_2015}. 
Moreover, strong evidence for splitting of Cooper pairs \cite{Hofstetter_2011,Heiblum_2012,Schindele_2012,Tarucha_2015}, which we shall link to the presence of odd-frequency spin-singlet $s$-wave pairing, has been presented. 
Here, we consider a system with two quantum levels at positions $\epsilon_{L,R}$. In the limit of large level separation at the quantum dots, it describes very well a DQD close to the crossing point of the dot resonances \cite{Burset_2011b}. 
In the absence of external magnetic fields and spin-orbit coupling terms, we describe the system in the combined Nambu-dot space using spinor fields $\Psi=(d_{L\up},d_{R\up},d^{\dagger}_{L\dw},d^{\dagger}_{R\dw})^T$, where $d_{\mu\sigma}$ ($d^{\dagger}_{\mu\sigma}$) annihilates (creates) an electron with spin $\sigma=\up,\dw$ at dot $\mu=L,R$. In the following, $\hat{\sigma}_{\nu}$ ($\hat{\tau}_{\nu}$), with $\nu=0,1,2,3$, are Pauli matrices in dot (Nambu) space, with identity matrix $\si$ ($\ti$). 
The Hamiltonian of the isolated DQD is given by 
$\check{H}_d\!=\! (\epsilon_L \hat{\sigma}_+ \!+\! \epsilon_R \hat{\sigma}_- \!+\! \Gamma_{LR}\sx)\tz$, 
with $\hat{\sigma}_\pm\!=\!(\si\pm\sz)/2$ and the inter-dot tunneling rate $\Gamma_{LR}>0$. 
Transport properties are characterized by the Green's function
\begin{equation}\label{eq:GF_DQD}
\check{g}(\omega)=\left[\omega\si\ti - \check{H}_d - \check{\Sigma}_N(\omega) - \check{\Sigma}_S(\omega) \right]^{-1} \, ,
\end{equation}
where $\omega$ denotes $\omega\!\pm\! i0^+$ or $i\omega_n$ for retarded/advanced or Matsubara Green's function, respectively, with $\omega_n\!=\!\pi(2n\!+\!1)k_BT$ for temperature $T$, Boltzmann constant $k_B$ and integer $n$. 
Following the geometry described in \fref{fig:figure_sketch}(a), we include the normal and superconducting leads as self-energies, 
\begin{align}\label{eq:se_N}
\check{\Sigma}_{N}(\omega)={}& is(\omega)(\Gamma_L \hat{\sigma}_+ + \Gamma_R \hat{\sigma}_- )\ti \, , \\
\label{eq:se_S}
\check{\Sigma}_{S}(\omega)={}& i(\Gamma_{SL} \hat{\sigma}_+ + \Gamma_{SR} \hat{\sigma}_-  ) [g(\omega)\ti-f(\omega)\tx] \, , 
\end{align}
with $s(\omega\!\pm\! i0^{+})\!=\!\mp1$, $s(i\omega_n)\!=\!\sgn(\omega_n)$, and $\Gamma_{\mu},\Gamma_{S\mu}\!>\!0$ the tunneling rates between dots and leads. 
We consider the regime where Kondo and exchange interactions between dots can be neglected. The effect of Coulomb repulsion on each dot is to renormalize the level positions $\epsilon_{\mu}$ and tunneling rares $\Gamma_{S\mu}$ \cite{Cuevas_2001,Burset_2011b}. 
We assume that the superconducting region is well described by a constant pair potential $\Delta$ and neglect its spatial dependence at the surface of the superconductor. The dimensionless Green's functions at the edge of the superconducting lead, for a BCS superconductor, are $f(\omega)\!=\!-(\Delta/\omega)g(\omega)\!=\!-\Delta/\sqrt{\omega^2-\Delta^2}$. 


\begin{table}
	\caption{\label{tab:sym} Symmetry classification of Cooper pairs according to frequency/spin/momentum/dot. From left to right, pairs can be even (E, $+$) or odd (O, $-$) under time-reversal (frequency), spin ($\sigma,\sigma'=\up,\dw$), momentum ($\mbf{k}$), and dot index (L, R), with S for spin singlet (top rows) and T for triplet (bottom rows). The last column shows the corresponding element of the anomalous Green's function: $F_{\mu\mu}$, with $\mu=L,R$, for local elements and $F_{s,a}$ for the nonlocal ones. }
	\begin{center}
		\begin{ruledtabular}
			\begin{tabular}{ c | c c c c | c }
				class & $\omega_n \rightarrow-\omega_n$ & $\sigma\leftrightarrow\sigma'$  &  $\mbf{k}\rightarrow-\mbf{k}$ & $L\leftrightarrow R$ & element \\ \hline
				\textcolor{red}{E}\textcolor{black}{S}\textcolor{black}{E}\textcolor{red}{E}  & \textcolor{red}{$+$} & \textcolor{black}{$-$} & \textcolor{black}{$+$} & \textcolor{red}{$+$} & $F_{\mu\mu}$, $F_{s}$ \\
				\textcolor{red}{O}\textcolor{black}{S}\textcolor{black}{E}\textcolor{red}{O}  & \textcolor{red}{$-$} & \textcolor{black}{$-$} & \textcolor{black}{$+$} & \textcolor{red}{$-$} & $F_{a}$ \\ \hline
				\textcolor{red}{E}\textcolor{black}{T}\textcolor{black}{E}\textcolor{red}{O} & \textcolor{red}{$+$} & \textcolor{black}{$+$} & \textcolor{black}{$+$} & \textcolor{red}{$-$} & $F_{a}$ \\
				\textcolor{red}{O}\textcolor{black}{T}\textcolor{black}{E}\textcolor{red}{E}  & \textcolor{red}{$-$} & \textcolor{black}{$+$} & \textcolor{black}{$+$} & \textcolor{red}{$+$} & $F_{\mu\mu}, F_{s}$ \\
			\end{tabular}
		\end{ruledtabular}
	\end{center}
\end{table}

\emph{Symmetry of induced pair amplitude.---} 
The uncoupled superconducting lead in \eref{eq:se_S} represents an even-frequency spin-singlet $s$-wave superconductor which satisfies $f(\omega_n) \!\!=\!\! f(-\omega_n)$, for Matsubara frequency. 
We analyze the symmetry of proximity induced pair amplitude in the DQD system from the anomalous part of the Green's function of \eref{eq:GF_DQD}, 
$F_{\mu\nu} \!=\! (\check{g})^{eh}_{\mu\nu}\!\sim\! \langle d_{\mu\up}d_{\nu\dw}\rangle$, 
with indexes in dot space $\mu,\nu\!=\! L,R$. 
Induced superconductivity in the DQD system thus acquires an extra orbital quantum number. Owing to this symmetry, the elements of $F_{\mu\nu}$ are divided into even- and odd-orbital terms. Defining $F_{s,a} \!=\! ( F_{LR} \!\pm\! F_{RL} )/2$, $F_{a}$ is the only element with odd parity in dot orbital degree of freedom. To be consistent with Fermi-Dirac statistics, $F_{a}$ must be odd in frequency. 
Explicitly, we find \footnotemark[1]
\begin{align}
		F_{s}(\omega_n)\!=\!{}& -i \frac{f(\omega_n)}{D(\omega_n)} \Gamma_{LR} \left( \Gamma_{SR} \epsilon_L + \Gamma_{SL} \epsilon_R \right) \, , \label{eq:agfs}\\
		F_{a}(\omega_n) \!=\!{}& \sgn(\omega_n) \frac{f(\omega_n)}{D(\omega_n)} \Gamma_{LR}  \left( \Gamma_{SR} \omega_L - \Gamma_{SL} \omega_R \right) \, , \label{eq:agfa}
\end{align}
with $\omega_{\mu} \!=\! |\omega_n| \!-\! \Gamma_{\mu}$ and $D(\omega_n) \!=\! \det[\check{g}^{-1}(\omega_n)] \!=\! D(-\omega_n)$. 
If $f(\omega_n)\!=\!f(-\omega_n)$ is satisfied, we find that $F_{s}(\omega_n) \!=\! F_{s}(-\omega_n)$ and $F_{a}(\omega_n) \!=\! - F_{a}(-\omega_n)$. 
A complete description of the allowed symmetries in the DQD system is given in \tref{tab:sym}, for both spin-singlet and triplet superconductors. 

\footnotetext[1]{See Appendix for more details on the definition of differential conductance, analytic expressions for the anomalous Green's function, a detailed description of the symmetry of induced pair amplitude, and results for the case with only one dot coupled to the superconducting lead. }

It is possible to enhance odd-frequency over even-frequency pairing on one of the dots, as it is sketched in \fref{fig:figure_sketch}(c). To study this effect, we define the local ratios \footnotemark[2]
\begin{equation}\label{eq:ratio-oe-LR}
R_{L,R}(\omega) = \frac{|F_{a}(\omega)|}{ \sqrt{ |F_{LL,RR}(\omega)|^2} + |F_{s}(\omega)|^2 }  \, .
\end{equation}
From Eqs. \ref{eq:agfs} and \ref{eq:agfa}, we see that nonlocal pair amplitudes are proportional to the inter-dot coupling $\Gamma_{LR}$, which is an essential element of our model \cite{[{Our symmetry analysis and conclusions are trivially extended to a recently proposed model for the DQD without inter-dot coupling which includes a third site representing the superconducting lead. }][{}]Dominguez_2015}. 
Moreover, $F_{s}$ is zero when the dot levels are on resonance ($\epsilon_{L}\!=\!\epsilon_{R}\!=\!0$). 
Therefore, dominant odd-frequency pairing on one of the dots requires left-right asymmetry, which can be achieved setting $\Gamma_{L}\!\neq\!\Gamma_{R}$ or $\Gamma_{SL}\!\neq\!\Gamma_{SR}$. Odd-frequency pairing is suppressed when the DQD levels are out of resonance ($\epsilon_{L}\!\neq\!0$ and/or $\epsilon_{R}\!\neq\!0$). 

\footnotetext[2]{It is possible to use real frequencies in \eref{eq:ratio-oe-LR} instead of Matsubara frequencies. In order to do so, one must define a Green's function with the same symmetry in both real and imaginary parts as $\check{g}(\omega_n)$ has on Matsubara frequency $\omega_n$ (see more details in the Supplementary information). }

\begin{figure}
	\includegraphics[width=1.\columnwidth]{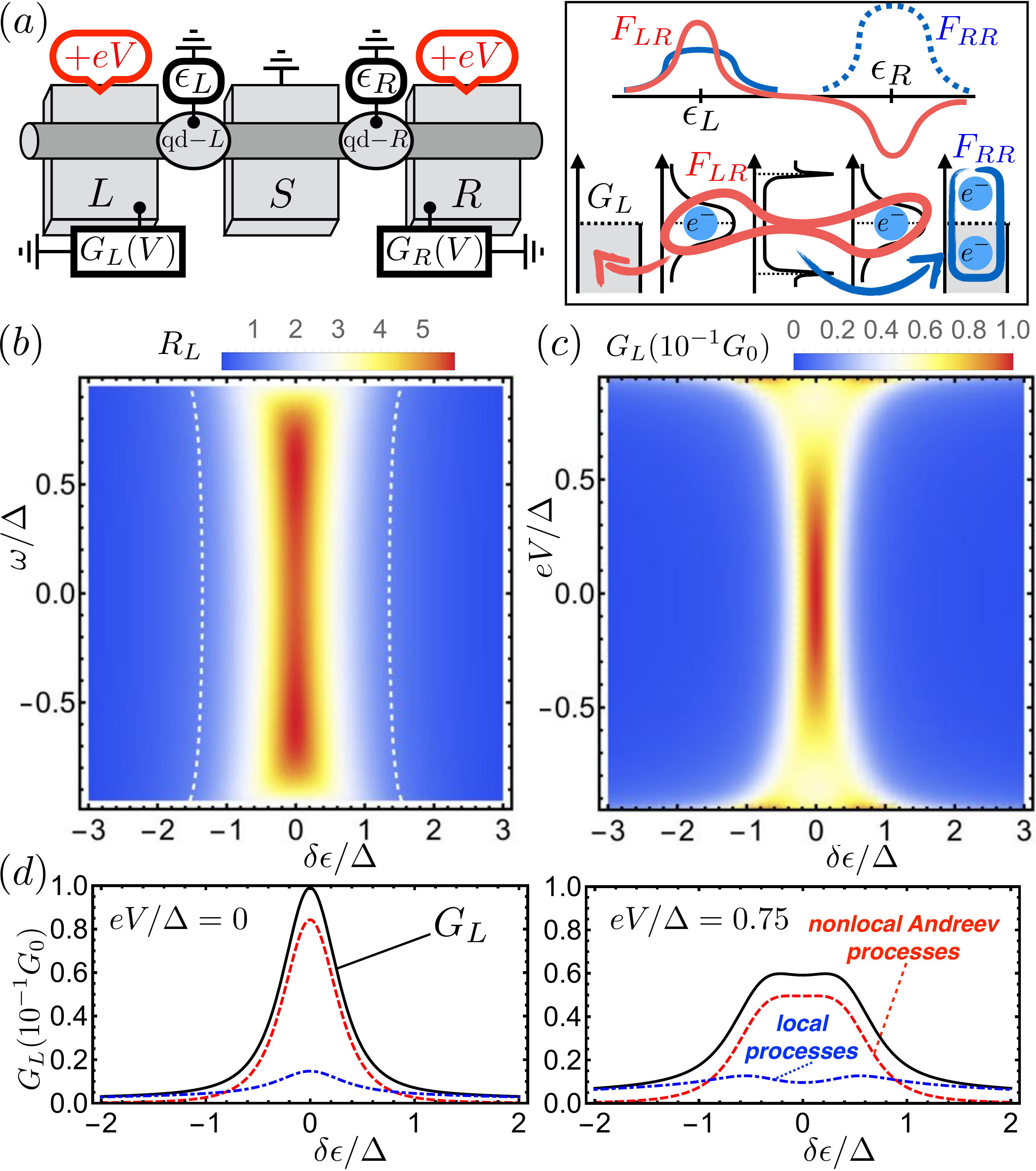}
	\caption{\label{fig:condL} 
		Cooper pair splitter configuration. 
		(a) A voltage $V$ is applied symmetrically to both normal electrodes which allows us to measure the conductances $G_{L}(V)$ and $G_{R}(V)$ (left). Right: In an asymmetric DQD system, odd-frequency nonlocal $F_{LR}$ (red line) can be enhanced on dot L and $G_L$ is mainly due to nonlocal Andreev processes (red arrow). 
		(b,c) Map of the ratio $R_L$ (b) and conductance $G_L$ (c) as a function of applied voltage and level position $\delta\epsilon=\epsilon_L+3\epsilon_R$. $R_L>1$ inside the white dashed line. 
		(d) Conductance (black solid lines), nonlocal Andreev (red dashed lines), and local processes (blue dot-dashed lines) for $eV=0$ (left) and $0.75\Delta$ (right). For all plots, $T=0$, $\Gamma_L/\Delta=5$, $\Gamma_R=\Delta=1$, $\Gamma_{SL}/\Delta=0.1$, $\Gamma_{SR}/\Delta=0.9$, and $\Gamma_{LR}/\Delta=0.5$. }
\end{figure}

\emph{Detection and manipulation of odd-frequency pairing.---} 
We consider two different transport measurements. 
First, a voltage bias $V$ is applied symmetrically to both normal leads [\fref{fig:condL}(a)]. This configuration is known as \textit{Cooper pair splitter} setup and has been used in recent experiments \cite{Takis_2010,Burset_2012b,Heiblum_2012,Schindele_2012}. 
At zero temperature, conductance at lead L is given by 
\begin{gather}\label{eq:DQD-G1}
G_{L}(V)= 2G_0 \left[T^{qp}_{L}(eV) + T^{eh}_{LL}(eV) + T^{eh}_{LR}(eV) \right] \, ,
\end{gather}
with $G_0=2e^2/h$ and quasiparticle tunneling transmission $T^{qp}_{L}$. 
$T^{eh}_{LL}$ and $T^{eh}_{LR}$ are the contributions from local and nonlocal Andreev processes, respectively. 
For subgap voltages ($|eV|\!<\!\Delta$), conductance is mainly given by Andreev processes while the quasiparticle contribution is almost negligible. 

\begin{figure}
	\includegraphics[width=1.\columnwidth]{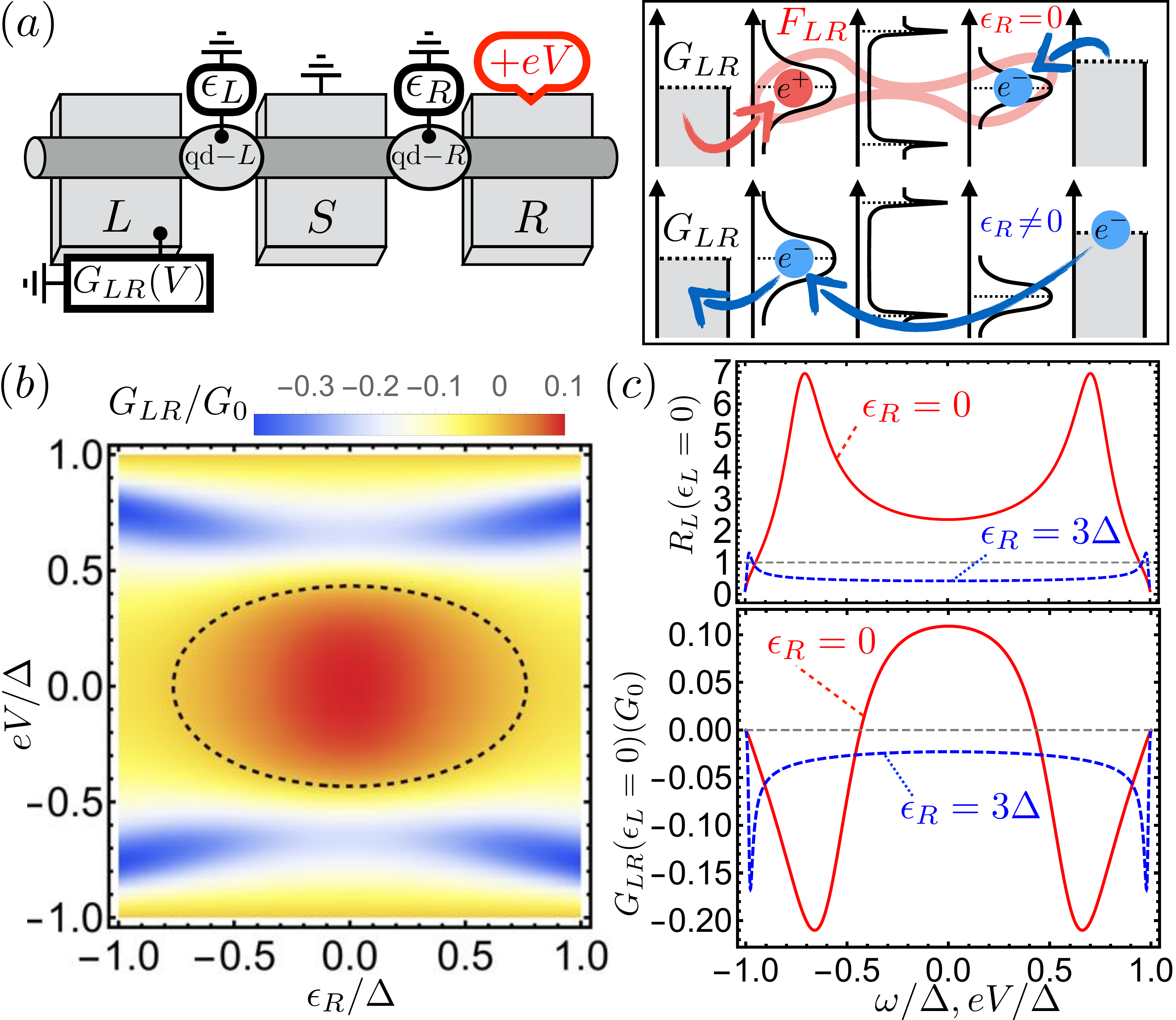}
	\caption{\label{fig:condLR} 
		Nonlocal conductance measurement. 
		(a) A voltage is applied to lead R allowing to measure the conductance $G_{LR}$ on lead L (left). Right: On resonance, odd-frequency nonlocal Cooper pairs provide a dominant contribution to $G_{LR}$ (red arrow), while electron tunneling dominates out of resonance (blue arrow). 
		(b) Map of $G_{LR}$ as a function of the applied voltage and the position of dot R, $\epsilon_{R}$, for $\epsilon_{L}=0$. $G_{LR}>0$ inside the black dashed line. 
		(c) Ratio on dot L (top) and nonlocal conductance (bottom) for $\epsilon_{L}=\epsilon_{R}$ (red solid line) and $\epsilon_{L}\neq\epsilon_{R}$ (blue dashed line). 
		For all plots, $\Gamma_L/\Delta=5$, $\Gamma_R/\Delta=0.2$, $\Gamma_{SL}=\Gamma_{SR}=\Gamma_{LR}=\Delta=1$, and $T=0$. }
\end{figure}

Alternatively, a current can flow through lead L if a voltage is applied to lead R [\fref{fig:condLR}(a)]. This is the basis for a \textit{nonlocal conductance measurement} \cite{Hofstetter_2010,Hofstetter_2011,Hakonen_2015} which, at zero temperature, is given by 
\begin{gather}\label{eq:DQD-G2}
G_{LR}(V)= G_0 \left[ T^{eh}_{LR}(eV) - T^{ee}_{LR}(eV) \right] \, ,
\end{gather}
where $T^{ee}_{LR}$ represents an electron tunneling process. 
At zero temperature, transmission probabilities for each process are calculated from the retarded Green's function as 
$T^{\alpha\beta}_{\mu\nu}(\omega)\!=\!4\Gamma_{\mu}\Gamma_{\nu} [ |(\check{g}^r)^{\alpha\beta}_{\mu\nu}(\omega\!+\!i0^+)|^2 + |(\check{g}^r)^{\alpha\beta}_{\mu\nu}(-\omega\!+\!i0^+)|^2]$, with $\alpha,\beta=e,h$ and $\mu,\nu=L,R$ \footnotemark[1]. 
Specifically, the transmission probability for Andreev processes reduces to 
\begin{equation}\label{eq:trans-ar}
T^{eh}_{\mu\nu}(\omega)=4\Gamma_{\mu}\Gamma_{\nu} [ |F_{\mu\nu}(\omega)|^2 + |F_{\mu\nu}(-\omega)|^2] \, . 
\end{equation}
Consequently, we can connect each microscopic process to a symmetry class. Indeed, local even-frequency pair amplitudes $F_{\mu\mu}$ provide the probability amplitude for transmission of the two electrons of a Cooper pair into the same lead, i.e., a local Andreev process sketched in \fref{fig:figure_sketch}(b). 
On the other hand, nonlocal components $F_{LR,RL}$ account for the probability amplitude of a process where the electrons of a Cooper pair split into different leads: a nonlocal Andreev process [\fref{fig:figure_sketch}(c)]. 
If both dot levels are aligned to the chemical potential, i.e., when the DQD is \textit{on resonance} with $\epsilon_{L}\!=\!\epsilon_{R}=0$, nonlocal pair amplitudes $F_{LR,RL}$ are odd in frequency. 
The presence of odd-frequency pairing in the DQD and its connection to a specific microscopic process that has been successfully observed in recent experiments is one of the main conclusions of this work. 


In an ideal setup, we can choose to uncouple one of the dots from the superconductor setting $\Gamma_{SL}\!=\!0$. 
Local transmission through that dot, $F_{LL}$ is suppressed and \eref{eq:ratio-oe-LR} reduces to $R_L\!=\!1/R_R\!=\!\sqrt{\omega^2\!+\!\Gamma_L^2}/\Gamma_{LR}$. As a result, for subgap energies $|\omega|\!<\!\Delta$, odd-frequency pairing becomes dominant at dot R (L) if $\Gamma_{L}\!<\!\Gamma_{LR}$ ($\Gamma_{L}\!>\!\Gamma_{LR}$) is satisfied. 
Consequently, in the Cooper pair splitter configuration, the conductance at one of the leads, \eref{eq:DQD-G1}, can be completely dominated by nonlocal Andreev processes, which is a signature of the presence of odd-frequency superconductivity \footnotemark[1]. 
Decoupling one of the dots requires careful patterning of the DQD, similarly to recent experiments in graphene \cite{Hakonen_2015}. 
In many other experiments, however, DQD are constructed by electrical confinement from electrodes on quasi-one dimensional materials \cite{Hofstetter_2010,Takis_2010}, as sketched in \fref{fig:condL}(a). 
It is thus challenging to decouple one of the dots from the superconductor. 
Therefore, we consider $\Gamma_{SL}\!\neq\!0$ in the following. 
We start with strong left-right asymmetry by setting $\Gamma_L\!\neq\!\Gamma_{R}$ and $\Gamma_{SL}\!\neq\!\Gamma_{SR}$ at the same time. 
To exclude double occupancy on the dots, we work in the regime with $\Gamma_{L,R}\!>\!\Gamma_{SL,SR}$ where a single-particle description of transport at the DQD system is allowed \cite{Recher_2001}. 
In \fref{fig:condL}(b) we show the ratio on dot L, $R_L$, as a function of $\omega$ and  $\delta\epsilon\!=\!\epsilon_{L}\!+\!\alpha\epsilon_{R}$, with $\alpha$ a constant. 
In agreement with our previous analysis, odd-frequency pairing is dominant on dot L for subgap energies as long as the dot levels are close to the chemical potential, i.e., for $|\omega|,|\delta\epsilon|\!\lesssim\!\Delta$. 
At zero temperature, the applied voltage corresponds to the frequency $\omega$. The conductance at lead L is enhanced for the same bias voltage regime, as shown in \fref{fig:condL}(c). 
Detailed analysis shows that the conductance is mainly given by nonlocal Andreev processes which stem from induced odd-frequency pairing [red lines in \fref{fig:condL}(d)]. 

A small degree of asymmetry is experimentally inevitable. However, by setting $\Gamma_{SL}\!\sim\!\Gamma_{SR}$ in the previous results, the contribution from local Andreev processes is enhanced and becomes comparable to that of nonlocal processes, making it more difficult to establish a connection between conductance and odd-frequency pairing. 
For weakly asymmetric setups, with $\Gamma_{SL}\!\sim\!\Gamma_{SR}$, it is better to perform a \textit{nonlocal conductance measurement} where odd-frequency induced nonlocal Andreev processes only compete with electron tunneling processes [see \fref{fig:condLR}(a)]. 
In principle, the two contributions should cancel each other \cite{Feinberg_2000,Martin_2002}. 
In a DQD three terminal device, however, the relative position of the dot levels becomes very important to favor Andreev processes, since they mainly take place when the levels are aligned on resonance. 
For this condition, the pair amplitude $F_{LR}$ is odd in frequency. 
Therefore, a positive nonlocal conductance proves the presence of odd-frequency pair amplitude \cite{[{Nonlocal conductance has also been suggested to detect odd-frequency spin-triplet pairing at the edge of two-dimensional topological insulators. }][{}] Crepin_2015}. 
Setting $\epsilon_{L}=0$, we show in \fref{fig:condLR}(b) a map of $G_{LR}$ as a function of $eV$ and $\epsilon_{R}$. Within the black dashed line, conductance is positive, i.e., dominated by nonlocal Andreev processes. 
As $|eV|\!\sim\!\Delta$, however, electron tunneling processes become more important and the conductance changes sign. 
The connection between positive Andreev-dominated conductance and odd-frequency symmetry is explicitly shown in \fref{fig:condLR}(c). 
When the dot R is on resonance (red solid lines), the nonlocal conductance is positive for subgap energies (bottom) while odd-frequency pair amplitude is dominant on dot L (top). If the dot R is taken out of resonance, the conductance becomes negative and odd-frequency pair amplitude is suppressed. 


\emph{Spin-triplet superconducting lead.---} 
When the central superconducting lead is a one-dimensional spin-triplet $p$-wave superconductor, like the case of a metallic nanowire on Sr$_2$RuO$_4$ \cite{Maeno_2012}, induced pairing amplitude at its edges is odd-frequency triplet $s$-wave represented by $f(\omega_n)\!=\!\Delta/\omega_n$, which displays Majorana edge states \cite{Asano_2013,Nakosai_2013,Egger_2016}. 
On the DQD, $F_{LL}$, $F_{RR}$, and $F_{s}$ are now odd-frequency functions, while $F_{a}$ is even-frequency pairing (see \tref{tab:sym}). 
Consequently, conductance measured in the Cooper pair splitter configuration is a very useful tool to study the symmetry of edge states at spin-triplet superconductors. 
For the perfectly symmetric case, where both $\Gamma_{L}\!=\!\Gamma_{R}\!\equiv\!\Gamma_{N}$ and $\Gamma_{SL}\!=\!\Gamma_{SR}\!\equiv\!\Gamma_{S}$ are satisfied, the even-frequency term $F_{a}$ vanishes and Cooper pairs injected on the same lead maintain the odd-frequency symmetry of the superconducting lead. 
$F_{a}$ increases proportionally to the difference between the tunneling rates for left and right dots. 
For example, if the asymmetry originates from the coupling to the normal leads (superconducting lead), even-frequency component follows $F_{a}\!\propto\!\Gamma_{S}\left(\Gamma_{L}\!-\!\Gamma_{R}\right)$ [$F_{a}\!\propto\!\left(\Gamma_{SL} \!-\! \Gamma_{SR}\right)\left(|\omega_n| \!-\! \Gamma_{N}\right)$]. 


\emph{Conclusions.---}
We propose a way to generate odd-frequency spin-singlet $s$-wave Cooper pairs on DQD-based three terminal devices. 
Due to the orbital degree of freedom in the DQD, symmetry of induced Cooper pairs can be broken, featuring a superposition of even and odd-frequency terms. 
Each symmetry type, however, is responsible for a different transport process; a feature unique of this setup. 
For spin-singlet superconductors, nonlocal Andreev processes on resonance are uniquely caused by odd-frequency pairing. 
Therefore, odd-frequency pairs can be detected from standard conductance measurements in asymmetric devices where the contribution of nonlocal Andreev processes is greatly enhanced. 
Additionally, it is possible to manipulate the amplitude of odd-frequency Cooper pairs by tuning the position of the dot levels. 
The situation is reversed if the central electrode is a spin-triplet $p$-wave superconductor. Odd-frequency triplet $s$-wave pairing is now associated to local Andreev processes which are the dominant contribution to the conductance if the dots are symmetrically coupled to the leads. 


\emph{Acknowledgments.---} 
We thank A. Levy Yeyati, N. Nagaosa, and B. Trauzettel for fruitful discussions and comments. 
We acknowledge financial support by JSPS International Research Fellowship (P.B.). 
This work was also supported by a Grant-in Aid for Scientific
Research on Innovative Areas “Topological Material Science”
(Grant Nos. 15H05853 and 15H05852), a Grant-in-Aid for Scientific Research
B (Grant Nos. 15H03686, 26287069, and 15H03525). 


\appendix

\section{Appendix A. Transport observables}
The Hamiltonian describing an isolated double-dot system reads 
\begin{gather*}
 H_{dqd}= \sum\limits_{\mu,\sigma} \tilde{\epsilon}_{\mu}n_{\mu\sigma} + \sum\limits_{\mu}U_{\mu} n_{\mu\up}n_{\mu\dw}  \nonumber \\ + \sum\limits_{\sigma} \Gamma_{LR} ( d^{\dagger}_{L\sigma} d_{R\sigma} + d^{\dagger}_{R\sigma} d_{L\sigma} ) \, ,
\end{gather*}
with $n_{\mu\sigma}= d^{\dagger}_{\mu\sigma} d_{\mu\sigma}$ for $\mu=L,R$ and $\sigma=\up,\dw$, where $d_{\mu\sigma}$ ($d^{\dagger}_{\mu\sigma}$) is the dot electron annihilation (creation) operator, $U_{\mu}$ the charging energy, $\tilde{\epsilon}_{\mu}$ the energy level, and $\Gamma_{LR}$ the hybridization between dots. 
We model the leads as non-interacting Fermi liquids with Hamiltonian
\begin{gather*}
H_{\mathrm{leads}}= \sum\limits_{\mu,\mbf{k},\sigma}\xi_{\mu,\mbf{k}} c^{\dagger}_{\mu,\mbf{k}\sigma} c_{\mu,\mbf{k}\sigma} \nonumber \\ 
+ \delta_{\mu,S} \frac{1}{2}\sum\limits_{\mu,\mbf{k},\sigma,\sigma'} \left( \Delta_{\mbf{k},\sigma\sigma'} c^{\dagger}_{S,\mbf{k}\sigma} c^{\dagger}_{S,-\mbf{k}\sigma'} + \mathrm{h.c.} \right)  \, ,
\end{gather*}
with $c_{\mu,\mbf{k}\sigma}$ ($c^{\dagger}_{\mu,\mbf{k}\sigma}$) the annihilation (creation) operator for electrons on lead $\mu=L,R,S$ with spin $\sigma=\up,\dw$, $\Delta_{\mbf{k},\sigma\sigma'}$ the pair potential and $\delta_{\mu,S}=1$ for $\mu=S$ and zero otherwise. 
For a conventional superconductor, pair potential is isotropic in momentum space (we assume it constant) while it depends on the momentum for triplet superconductor. In our effective model, quantum dots are considered almost point-like so electron fields are integrated out and the magnitude of the pair potential for triplet superconductor can also be considered constant. We choose $\Delta\hat{s}_y$ for spin-singlet superconductors and $\Delta\hat{s}_x$ for spin-triplet ones, with $\Delta>0$ constant, and $\hat{s}_{x,y,z}$ Pauli matrices acting in spin space. Consequently, the system is spin degenerate for both types of superconductors and we work only in Nambu-dot space. 
Finally, the tunneling Hamiltonian between dots and leads reads as
\begin{gather*}
H_{\mathrm{tunnel}}= \sum\limits_{\mu,\mbf{k},\sigma} \left( t_{\mu} d^{\dagger}_{\mu\sigma} c_{\mu,\mbf{k}\sigma} + t_{S\mu} d^{\dagger}_{\mu\sigma} c_{S,\mbf{k}\sigma} + \mathrm{h.c.} \right)  \, ,
\end{gather*}
for $\mu=L,R$ and $t_{\mu,S\mu}$ the tunneling amplitudes. We define the tunneling rates as $\Gamma_{\mu}=\pi t_{\mu}^2\rho_{\mu}$ for the normal leads and $\Gamma_{S\mu}=\pi t_{S\mu}^2\rho_{S}$ for the superconducting lead, with $\rho_{\mu}$ and $\rho_{S}$ the normal state density of states at the Fermi level of normal and superconducting leads, respectively. 

In the following, we consider the regime where Kondo correlations can be neglected and where both dots are close to resonance. In this limit, within Hartree-Fock approximation, Coulomb interactions on each dot give rise to the self-energies $\hat{\Sigma}^{r}_{U\!\mu}\!=\!U_{\mu}(\langle n_{\mu}\rangle \tz \!+\! \langle d^{\dagger}_{\mu\up} d^{\dagger}_{\mu\dw} \rangle \tx)$. Using the equation of motion technique, the interacting Green's function reduces to the non-interacting one with renormalized couplings $\Gamma_{S\!\mu}\!=\!\tilde{\Gamma}_{S\mu} \!-\! U_{\mu}\langle d^{\dagger}_{\mu\up} d^{\dagger}_{\mu\dw} \rangle$ and level positions $\epsilon_{\mu}\!=\! \tilde{\epsilon}_{\mu} \!+\! U_{\mu}\langle n_{\mu}\rangle$, where $\tilde{\Gamma}_{S\mu}$ and $\tilde{\epsilon}_{\mu}$ are the bare coupling and level energies \cite{Cuevas_2001,Burset_2011b}. 


\begin{figure}
	\includegraphics[width=1.\columnwidth]{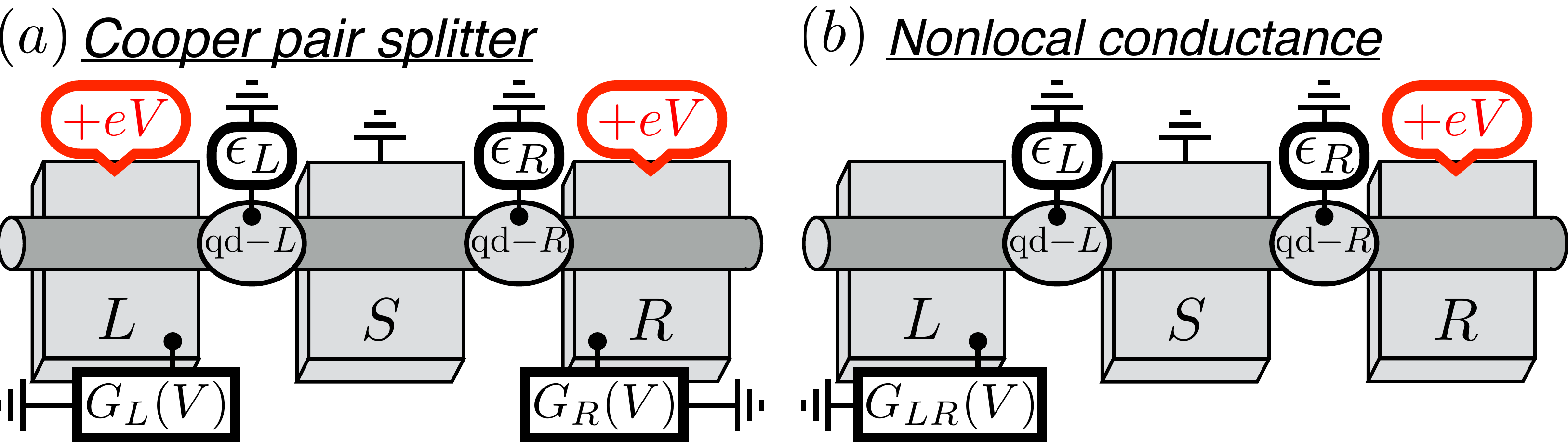}
	\caption{\label{fig:cond-mes} 
		Conductance measurements in a DQD three-terminal device. 
		Schematics of DQD geometry where a quasi-one dimensional system is contacted by two normal leads L, R and a central superconducting lead S. Quantum dots are formed by electric confinement between leads and can be independently controlled by side gates $\epsilon_{L,R}$. (a) Cooper pair splitter configuration where the normal leads are equally biased. A current flows from S to the normal leads where we can measure the conductances $G_{L}$ and $G_{R}$. (b) Nonlocal conductance measurements require that only lead R is biased. A current can still flow to lead L, generating nonlocal conductance $G_{LR}$. 
	}
\end{figure}


The steady state current between dot and lead $\mu=L,R$ is defined as \cite{Meir_1992}
\begin{align*}
 I_{\mu} = {}& i\frac{e}{\hbar} \sum\limits_{\mbf{k}\sigma}t_{\mu} \left( \langle c^{\dagger}_{\mu,\mbf{k}\sigma}d_{\mu\sigma} \rangle  - \langle d^{\dagger}_{\mu\sigma}c_{\mu,\mbf{k}\sigma} \rangle\right) \nonumber \\
 = {}& \frac{e}{\hbar} \int \frac{\mathrm{d}\omega}{2\pi} \mathrm{Tr}\left\{ \check{V}_{\mu} \left[ \check{G}^{<}_{d\mu}(\omega) - \check{G}^{<}_{\mu d}(\omega) \right] (\si\tz) \right\} \, ,
\end{align*}
with $\check{V}_{L,R}=t_{L,R}\hat{\sigma}_{\pm}\tz$ and $\hat{\sigma}_{\pm}\!=\!(\si\!\pm\!\sz)/2$. We have used the Fourier transform of the Keldysh lesser Green's functions $\check{G}^{<}_{\mu d}(t-t')\!=\!-i\mean{ c^{\dagger}_{\mu,\mbf{k}\sigma}(t') d_{\mu\sigma}(t) }$ and $\check{G}^{<}_{d\mu}(t-t')\!=\!-i\mean{ d^{\dagger}_{\mu\sigma}(t') c_{\mu,\mbf{k}\sigma}(t) }$. All matrices are defined in the Nambu-dot space used in the main text. 
Following Refs.~\onlinecite{Cuevas_1996,Sun_1999}, the current expression can be written in terms of the retarded Green's function of the DQD, $\check{g}^{r}$, defined in the main text. For the sake of completeness, we explicitly write the Green's function in dot space, namely, 
\begin{widetext}
\begin{equation*}
\check{g}^{r,a}(\omega) = \left(\! \begin{array}{cc} \!\left[ \omega\!-\!is(\omega)\Gamma_{L}\!-\!i\Gamma_{SL}g(\omega) \right]\ti\!-\!\epsilon_{L}\tz \!+\!i\Gamma_{SL} f(\omega) \tx & -\Gamma_{LR}\tz \\ -\Gamma_{LR}\tz  & \!\left[ \omega\!-\!is(\omega)\Gamma_{R}\!-\!i\Gamma_{SR}g(\omega) \right]\ti\!-\!\epsilon_{R}\tz \!+\!i\Gamma_{SR} f(\omega)\tz \end{array}\!\right)^{-1} \, ,
\end{equation*}
where $\omega$ stands for $\omega\!\pm\! i0^+$ and we have defined $s(\omega\!\pm\!i0^+)\!=\!\mp1$. The Pauli matrices $\hat{\tau}_{1,2,3}$ and identity matrix $\ti$ act in Nambu space. 
\end{widetext}
For subgap voltages (i.e. when $|\omega|<\Delta$), the current at normal electrodes is given by
\begin{align*}
I_{\mu} ={}& \frac{2e}{h} \int\mathrm{d}\omega \left[ \left( f^{e}_{\mu} - f^{h}_{\mu} \right) T^{eh}_{\mu\mu} \right. \\ & \left. + \left( f^{e}_{\mu} - f^{h}_{\bar{\mu}} \right) T^{eh}_{\mu\bar{\mu}} + \left( f^{e}_{\mu} - f^{e}_{\bar{\mu}} \right) T^{ee}_{\mu\bar{\mu}} \right] \, ,
\end{align*}
where if $\mu=L,R$ then $\bar{\mu}=R,L$ and with $f_{\mu}^{e,h}=1/(1+\exp\left[(\omega\pm eV_{\mu})/(k_BT)\right])$. All the microscopic processes that give a contribution to the subgap current are summarized in this important result. Indeed, the current measured at electrode L depends on the local Andreev reflection that takes place at that lead, namely, 
\begin{equation*}
T^{eh}_{LL}(\omega) = 8\Gamma_L^2 |(\check{g}^r)^{eh}_{LL}(\omega)|^2 \, ,
\end{equation*}
where we have used that $(\check{g})^{eh}_{\mu\mu}(\omega)\!=\!(\check{g})^{he}_{\mu\mu}(\omega)$ and $(\check{g})^{\alpha\bar{\alpha}}_{\mu\mu}(\omega)\!=\!(\check{g})^{\alpha\bar{\alpha}}_{\mu\mu}(-\omega)$, with $\bar{\alpha}\!=\!h,e$ for $\alpha\!=\!e,h$ and $\mu\!=\!L,R$ labeling Nambu and dot space, respectively. 
Also included in the current are nonlocal Andreev processes
\begin{align*}
T^{eh}_{LR}(\omega) ={} &  4\Gamma_L \Gamma_R \left[ |(\check{g}^r)^{eh}_{LR}(\omega)|^2 + |(\check{g}^r)^{eh}_{LR}(-\omega)|^2 \right] \, ,
\end{align*}
and electron tunneling processes
\begin{equation*}
T^{ee}_{LR}(\omega) = 4\Gamma_L\Gamma_R \left[|(\check{g}^r)^{ee}_{LR}(\omega)|^2 + |(\check{g}^r)^{ee}_{LR}(-\omega)|^2 \right] \, .
\end{equation*}

On the other hand, for $|\omega|>\Delta$, the quasiparticle contribution to the current is 
\begin{align*}
I_{L} ={}& \frac{2e}{h} \!\int\!\!\mathrm{d}\omega \left[ \left( f_{S} - f^{e}_{\mu} \right) T^{qp,e}_{L}  - \left( f_{S} - f^{h}_{\mu} \right) T^{qp,h}_{L} \right] \, , \\
T^{qp,\alpha}_{L}(\omega) ={}& 4\Gamma_L \real\left\{ \left[ \Gamma_{SL} \hat{G}_{LL}^{\dagger}(\omega) \hat{g}_{S}(\omega) \hat{G}_{LL}(\omega) \right. \right. \\ & \left.\left. + \Gamma_{SR} \hat{G}_{RL}^{\dagger}(\omega) \hat{g}_{S}(\omega) \hat{G}_{LR}(\omega)  \right]_{\alpha\alpha} \right\} \, , \\
\hat{g}_{S}(\omega) ={}& g(\omega)\ti - f(\omega) \tx \, ,
\end{align*}
for $\alpha=e,h$,  $f_{S}=1/(1+\exp\left[\omega/(k_BT)\right])$, and with Nambu matrices $(\hat{G}_{\mu\nu})^{\alpha\beta}=(\check{g}^{r})^{\alpha\beta}_{\mu\nu}$. 

In the Cooper pair splitter setup, we apply a voltage difference between the superconductor and both normal leads. 
Differential conductance is thus calculated simultaneously biasing the normal leads with $V_L=V_R\equiv V$. At zero temperature, it reads
\begin{align}\label{eq:DQD-GL1}
G_{L}(V)\equiv \frac{\partial I_L}{\partial V} ={}& 2G_0 \left[ T^{qp}_{L}(eV) + T^{eh}_{LL}(eV) + T^{eh}_{LR}(eV) \right] \, , \nonumber 
\end{align}
with $G_0=2e^2/h$ and $T^{qp}_{L}(eV)=T^{qp,e}_{L} (eV) + T^{qp,e}_{L} (-eV)$. 

On the other hand, in a nonlocal conductance measurement, the current flowing through lead L originates from a voltage drop applied only to lead R. Therefore, setting $V_L=0$, at zero temperature, we find
\begin{equation}\label{eq:DQD-GL2}
G_{LR}(V)\equiv \frac{\partial I_L}{\partial V_R} = G_0 \left[ T^{eh}_{LR}(eV) - T^{ee}_{LR}(eV) \right] \, . \nonumber 
\end{equation}
$G_{LR}$ is only given by electron tunneling and nonlocal Andreev processes and is positive when the latter are dominant. 


\section{Appendix B. Anomalous Green's function}
We define the anomalous Green's function $F_{\mu\nu}(\omega) =(\check{g})^{eh}_{\mu\nu}(\omega)$. By choosing Matsubara frequencies, we can analyze each component, namely, 
\begin{widetext}
	\begin{subequations}\label{eq:f-comps}
		\begin{align}
		F_{LL}(\omega_n)={}& i \left[ \epsilon_R^2\Gamma_{SL} + \Gamma_{SR}\Gamma_{LR}^2+ \Gamma_{SL} \left( \omega_R^2 +\Gamma_{SR}^2 -2|g|\Gamma_{SR}\omega_R \right)  \right] \frac{f(\omega_n)}{D(\omega_n)} \, , \\
		F_{RR}(\omega_n)={}& i \left[ \epsilon_L^2\Gamma_{SR} + \Gamma_{SL}\Gamma_{LR}^2+ \Gamma_{SR} \left( \omega_L^2 +\Gamma_{SL}^2 -2|g|\Gamma_{SL}\omega_L \right)  \right] \frac{f(\omega_n)}{D(\omega_n)} \, , \\
		F_{s}(\omega_n)={}& \frac{F_{LR}+F_{RL}}{2} = -i \frac{f(\omega_n)}{D(\omega_n)} \Gamma_{LR} \left( \epsilon_L\Gamma_{SR} + \epsilon_R\Gamma_{SL} \right) \, , \\
		F_{a}(\omega_n)={}& \frac{F_{LR}-F_{RL}}{2} = \sgn(\omega_n) \frac{f(\omega_n)}{D(\omega_n)} \Gamma_{LR}  \left[ \left( \Gamma_{SR} - \Gamma_{SL}\right)|\omega_n| + \Gamma_{SR}\Gamma_{L}-\Gamma_{SL}\Gamma_{R} \right] \, ,
		\end{align}
	\end{subequations}
\end{widetext}
where we have defined $\omega_{L,R}\equiv |\omega_n| - \Gamma_{L,R}$ and the denominator is given by
\begin{gather*}
D(\omega_n)=\Gamma_{LR}^4 + 2\Gamma_{LR}^2  \nonumber \\ \times \left[ \omega_L\omega_R + \Gamma_{SL}\Gamma_{SR} - |g|\left( \Gamma_{SR}\omega_L + \Gamma_{SL}\omega_R \right) - \epsilon_L\epsilon_R \right] \nonumber \\ 
+ \left( \omega_L^2 + \Gamma_{SL}^2 - 2|g|\Gamma_{SL}\omega_L \right) \left( \omega_R^2 + \Gamma_{SR}^2 - 2|g|\Gamma_{SR}\omega_R \right) \, . \nonumber
\end{gather*}
It is easy to check that the denominator fulfills $D(\omega_n)\!=\!D(-\omega_n)$. Therefore, the symmetry with respect to frequency is given by the numerator. 
Local terms $F_{LL}(\omega_n)$ and $F_{RR}(\omega_n)$ and the symmetric nonlocal term $F_{s}(\omega_n)$, which is zero if $\epsilon_L\!=\!\epsilon_R=0$, maintain the symmetry of the superconducting lead determined by $f(\omega_n)$. 
On the other hand, it is clear from \eref{eq:f-comps}(d) that the symmetry of the anti-symmetric nonlocal pairing $F_{a}(\omega_n)$ is determined by $\sgn(\omega_n) f(\omega_n)$. 
It represents the presence of induced odd-frequency spin-singlet $s$-wave pairing in the DQD system if $f(\omega_n)$ is an even function of $\omega_n$, as it is the case of BCS spin-singlet $s$-wave superconductors. 
Alternatively, $F_{a}(\omega_n)$ is an even-frequency function if $f(\omega_n)$ is odd, which is the symmetry of the edge states of a spin-triplet $p$-wave superconductor. 
$F_{a}(\omega_n)$ is finite only in the presence of inter-dot coupling ($\Gamma_{LR}\neq0$). It also vanishes if $\Gamma_{L}\!=\!\Gamma_{R}$ and $\Gamma_{SL}\!=\!\Gamma_{SR}$ are satisfied simultaneously. 

For retarded and advanced Green's functions, $\omega\!\rightarrow\! E \pm i0^{+}$, and the symmetry analysis is not straightforward. Due to the infinitesimal imaginary part, retarded and advanced Green's functions are complex functions with real and imaginary parts that have different dependence on the energy $E$. It is possible, however, to construct a Green's function that depends on real energy and has the same symmetry as the Matsubara Green's function \cite{Burset_2015}. Namely, 
\begin{gather}
F_{\mu\nu}[E+i\sgn(E)0^+] = \\ \Theta(-E)F^a_{\mu\nu}(E-i0^+)+\Theta(E)F^r_{\mu\nu}(E+i0^+) \, , \nonumber
\end{gather}
where $E$ is a real variable which we associate with the frequency $\omega$. $F_{\mu\nu}(E)$ is also a complex function and its real and imaginary parts have the same symmetry with $E$ as $F_{\mu\nu}(\omega_n)$ has with $\omega_n$. 


\section{Appendix C. Symmetry classification of pair amplitude}
Regarding its degrees of freedom, the pair amplitude can be even (E) or odd (O) with respect to the frequency, momentum, and dot orbital component (i.e., L or R). It can also be a spin-singlet (S) or triplet (T) component. 
Total symmetry is constrained to be antisymmetric under the exchange of all degrees of freedom.  
Following Ref. \onlinecite{Tanaka_JPSJ}, we adopt the convention \textit{frequency/spin/momentum/orbital}. In the effective model, momentum has been integrated out and all allowed symmetries are even in this quantity. 
We list all possible terms in \tref{tab:sym}. 
For the superconducting leads, the dot degree of freedom must also be even. 
We thus classify a conventional BCS superconductor, which is even in frequency and spin-singlet, as ESEE. 
On the other hand, superconductivity at the edge of a one-dimensional quantum wire on top of an unconventional triplet superconductor is odd in frequency \cite{Asano_2013}. It is thus classified as OTEE. The pair amplitude induced in the double-dot features two symmetries. 
The local and symmetric components $F_{\mu\mu,s}$ maintain the symmetry of the superconducting lead and are thus either ESEE or OTEE, depending on the spin state of the Cooper pairs in the superconductor. 
The antisymmetric component $F_{a}$, however, changes the orbital symmetry. Consequently, frequency dependence is also changed to OSEO or ETEO. 


\begin{figure}
	\includegraphics[width=1.\columnwidth]{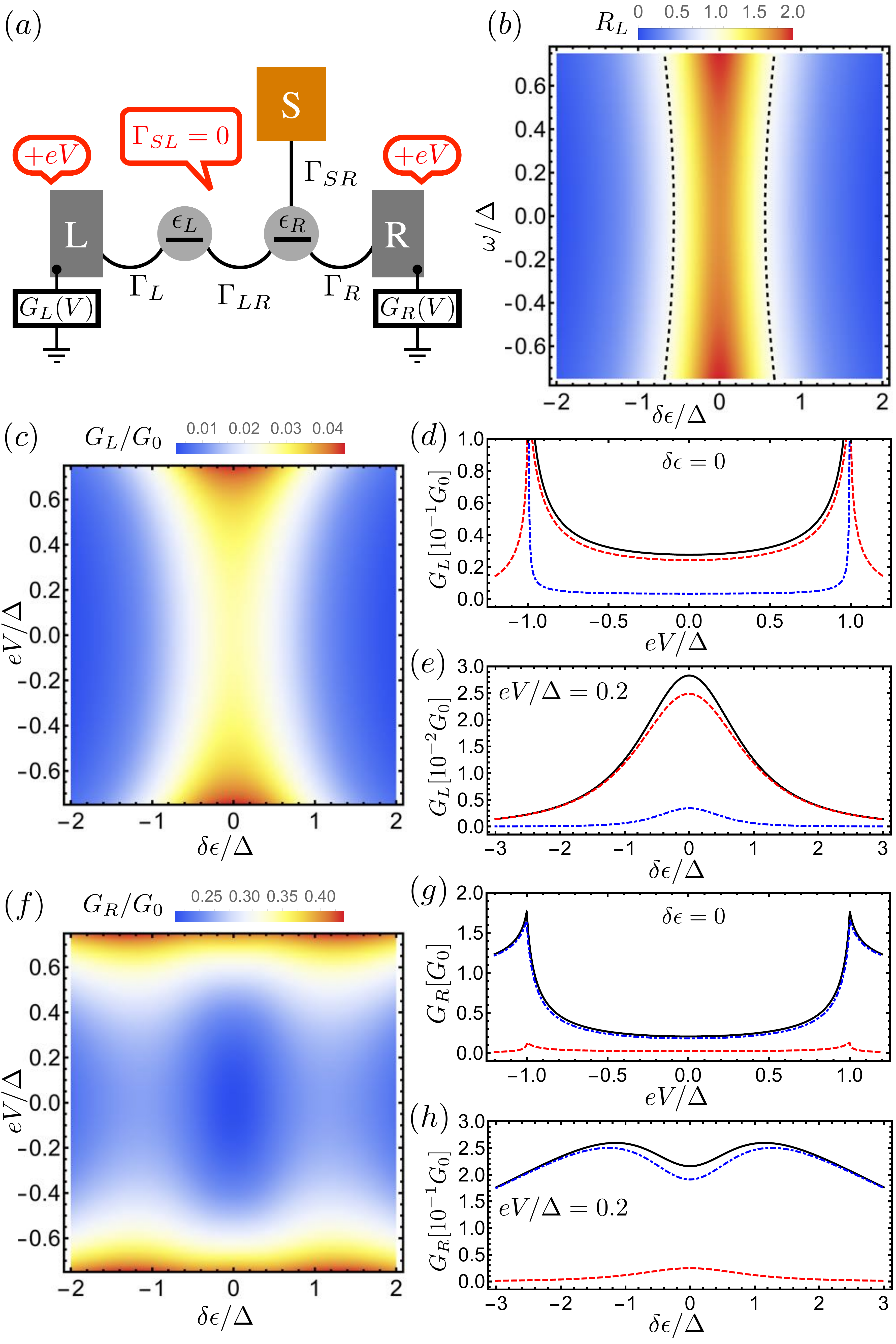}
	\caption{\label{fig:dqd-lateral}
		Conductance in the Cooper pair splitter configuration for the case with only one dot coupled to the superconductor. 
		(a) Sketch of the setup where the superconducting lead is only coupled to dot R ($\Gamma_{SL}=0$). (b) Ratio on dot L is greater than 1 within the black dashed line. (c) Conductance on dot L as a function of energy $\omega$ and dot levels $\delta\epsilon=2\epsilon_L+\epsilon_R$. (d,e) Conductance on lead L as a function of $\omega$ (d) or $\delta\epsilon$ (e) (black lines). Red lines correspond to nonlocal Andreev processes and blue line to local processes. (f) Map of conductance on lead R. (g,h) Conductance (black lines), nonlocal Andreev processes (red dashed lines), and local processes (blue dot-dashed lines) as a function of the $\omega$ (g) or $\delta\epsilon$ (h). For all plots, $\Gamma_{L}=1.5\Delta$, $\Gamma_{R}=5\Delta$, $\Gamma_{SR}=\Gamma_{LR}=\Delta=1$. 
	}
\end{figure}


\section{Appendix D. Superconductor coupled to one dot only}
In this section we consider the ideal case where only one of the dots is coupled to the superconducting lead. In the following, we assume $\Gamma_{SL}=0$ as it is sketched in \fref{fig:dqd-lateral}(a). We only consider the case of an even-frequency spin-singlet $s$-wave superconductor. 
By canceling the coupling between central lead and dot L, the ratios between odd- and even-frequency components become very simple, namely, 
\begin{align*}
 R_L(\omega_n>0) ={}& \frac{ \lvert\omega_n-\Gamma_{L}\rvert }{ \sqrt{\Gamma_{LR}^2+\epsilon_{L}^2} } \, , \\
 R_R(\omega_n>0) ={}& \frac{ \Gamma_{LR} \lvert\omega_n-\Gamma_{L}\rvert }{ \sqrt{ \Gamma_{LR}^2\epsilon_{L}^2 + \left( \left(\omega_n-\Gamma_{L}\right)^2 + \epsilon_{L}^2 \right)^2 } } \, .
\end{align*}
It is interesting to notice that the ratios do not depend on the parameters from dot R, namely, $\Gamma_{R}$, $\Gamma_{SR}$, and $\epsilon_{R}$. Additionally, when $\epsilon_{L}=0$, we find that $R_{L}=1/R_{R}=\lvert\omega_n-\Gamma_{L}\rvert/\Gamma_{LR}$. Consequently, odd-frequency becomes dominant on dot L and is suppressed on dot R if $\Gamma_{L}>\Gamma_{LR}$. 
Such behavior is reversed in the opposite regime with $\Gamma_{L}<\Gamma_{LR}$. 

We show in \fref{fig:dqd-lateral}(b) the ratio on dot L as a function of the energy and level position $\delta\epsilon=2\epsilon_{L}+\epsilon_{R}$ for $\Gamma_{LR}=\Delta=2\Gamma_{L}/3$. Odd-frequency pairing is dominant for subgap energies in the region $|\delta\epsilon|\lesssim\Delta$. 
We now study the conductance in the Cooper pair splitter configuration. We set $\Gamma_{R}=5\Gamma_{SR}$, with $\Gamma_{SR}=\Delta$ to allow for a single-particle description. The conductance at lead L and R are shown in \fref{fig:dqd-lateral}(c) and \fref{fig:dqd-lateral}(f), respectively. Their behavior for subgap energies looks very different, with $G_{L}$ featuring a strong contribution at $|\delta\epsilon|\lesssim\Delta$, while $G_{R}$ displays maxima around $|\delta\epsilon|\sim\Delta$. 
We analyze the contribution to the conductance from local and nonlocal Andreev processes in \fref{fig:dqd-lateral}(d,e) for $G_{L}$ and in \fref{fig:dqd-lateral}(g,h) for $G_{R}$. It is clear that nonlocal processes are dominant for $G_{L}$ (red dashed lines) while the local processes are the main contribution to $G_{R}$ (blue dot-dashed lines). 

As a result, in the extreme case where only one of the dots is connected to the superconducting lead, induced odd-frequency pairing can be maximized on the opposite dot. Consequently, an enhanced conductance that is mainly caused by nonlocal Andreev processes stemming from odd-frequency pair amplitude can be measured on that dot. 


%

\end{document}